\documentclass[twocolumn,preprintnumbers,amsmath,amssymb,prb,superscriptaddress]{revtex4}

\usepackage{graphicx}
\usepackage{dcolumn}
\usepackage{bm}
\usepackage{color}
\usepackage{epsfig}

\begin{document}

\newcommand{\commute}[2]{\left[#1,#2\right]}
\newcommand{\bra}[1]{\left\langle #1\right|}
\newcommand{\ket}[1]{\left|#1\right\rangle }
\newcommand{\anticommute}[2]{\left\{  #1,#2\right\}  }
\newcommand{\exsum}[2]{\substack{#1\neq #2}}
\renewcommand{\arraystretch}{2}
\title{Nuclear spin dynamics and Zeno effect in quantum dots and
  defect centers} 
 
\author{D. Klauser}
\affiliation{Department of Physics, University of Basel,
Klingelbergstrasse 82, CH-4056 Basel, Switzerland}
\author{W. A. Coish}
\affiliation{Department of Physics, University of Basel,
Klingelbergstrasse 82, CH-4056 Basel, Switzerland}
\affiliation{Institute for Quantum Computing and Department of Physics
and Astronomy, University of Waterloo, 200 University Ave. W.,
Waterloo, ON, N2L 3G1, Canada}
\author{Daniel Loss}
\affiliation{Department of Physics, University of Basel,
Klingelbergstrasse 82, CH-4056 Basel, Switzerland}
 
\date{\today}
\begin{abstract}

We analyze nuclear spin dynamics in quantum dots and defect centers
with a bound electron 
under electron-mediated coupling between nuclear spins due to the
hyperfine interaction (``J-coupling'' in NMR). Our analysis shows that 
the Overhauser field generated by the nuclei at the position of 
the electron has short-time dynamics quadratic in time for an initial 
nuclear spin state without transverse coherence. The quadratic
short-time behavior allows for an extension of the Overhauser field
lifetime through a sequence of projective measurements (quantum Zeno
effect). We analyze the requirements on the repetition rate of
measurements and the measurement accuracy to achieve such an
effect. Further, we calculate the long-time behavior of the Overhauser
field for effective electron Zeeman splittings larger than the
hyperfine coupling strength and find, both in a Dyson series expansion
and a generalized master equation approach, that for a nuclear spin
system with a sufficiently smooth polarization the electron-mediated
interaction alone leads only to a partial decay of the Overhauser
field by an amount on the order of the inverse number of nuclear spins
interacting with the electron.     

\end{abstract} 

\maketitle

\section{Introduction}\label{sec:introduction}
Technological advancements have made it possible to confine very few
electrons in a variety of nanostructures such as nanowires, quantum
dots, donor impurities, or defect 
centers.\cite{Tarucha:1996a,Miller:1997a,Ciorga:2000a,Kouwenhoven:2001a,Elzerman:2003a,Petta:2004a,Zumbuhl:2004a,Schleser:2004a,Mason:2004a,Jelezko:2004a,Petta:2005b,Jarillo:2005a,Koppens:2006a,Hanson:2006a,Childress:2006a,Graeber:2006a,Shorubalko:2007a,Fasth:2007a,Simmons:2007a,Hu:2007a,Joergensen:2007a,Hanson:2007a}
One driving force behind these achievements is a series of proposals
for using the spin of an electron as a qubit for quantum computing.
\cite{Loss:1998a,Imamoglu:1999a,Wrachtrup:2001a} This spin interacts
with the nuclear spins in the host material via the hyperfine
interaction. While this interaction leads to decoherence of the
electron spin state on one hand, it also provides the opportunity to
create a local effective magnetic field (Overhauser field) for the
electron by inducing polarization in the nuclear spin system, which
could be used, e.g., for rapid single-spin
rotations.\cite{Coish:2006b} Polarizing the nuclear spin system is
also one possible way to suppress hyperfine-induced decoherence
\cite{Coish:2004a} or it can be used as a source of spin polarization
to generate a spin-polarized current. In any case, 
controlling the dynamics of the Overhauser field and, in particular,
to prevent its decay, is thus of vital importance in the context of
spintronics and quantum computation.\cite{Awschalom:2002a}

In GaAs quantum dots the Overhauser field can become as large as $5$T. The
build-up, decay, and correlation time of the Overhauser field have been
studied in a number of
systems,\cite{Paget:1982a,Huettel:2003a,Bracker:2004a,Lai:2005a,Koppens:2005a,Maletinsky:2007a,Baugh:2007a,Reilly:2007a,Makhonin:2007a,Foletti:2008a,Christ:2007a,Rudner:2007a,Rudner:2007b,Danon:2008a} 
suggesting timescales for the decay on the order of seconds,
minutes, or in one case, even hours.\cite{Greilich:2007a} 

The dynamics of the Overhauser field are governed by the mutual
interaction between the nuclear spins. There is on one hand the direct
dipolar coupling between the nuclear spins. On the other hand, due
to the presence of a confined electron, there is also an indirect
interaction: The coupling of the nuclear spins to the electron via the
hyperfine interaction leads to an effective interaction between
the nuclear spins that is known as the electron-mediated interaction. 
While the effect of this electron-mediated
interaction on the decoherence of the \emph{electron} has been studied 
previously,\cite{Yao:2006a,Deng:2006a,Coish:2007a} theoretical
studies of the decay of the Overhauser field have so far studied
direct dipole-dipole interaction and the effect of the hyperfine
interaction was taken into account through the Knight shift that the
electron induces via the hyperfine interaction.\cite{Deng:2005a} In
this article we investigate the effect of the electron-mediated
interaction between nuclear spins on the dynamics of the Overhauser
field. While the direct dipolar coupling is always present, it can be
weaker than the electron-mediated interaction for
magnetic fields that are not too large and may be further reduced via NMR
pulse sequences or by diluting the concentration of nuclear
spins.\cite{diluting}  We find
in our calculation that, for effective electron Zeeman splittings
$\omega$ (sum of Zeeman splittings due to the external magnetic field
and Overhauser field) larger than the hyperfine coupling strength $A$, the
decay of the Overhauser field due to the electron-mediated interaction is
incomplete, i.e., that only a small fraction of the Overhauser field
decays. In a short-time expansion that is valid for $\omega$ larger
than $A/\sqrt{N}$, where $N$ is the number of nuclear spins with which
the electron interacts, we find a quadratic initial decay on a
timescale $\tau_e=N^{3/2}\omega/A^2$. We show that, by performing
repeated projective measurements on the Overhauser field, a quantum Zeno
effect occurs, which allows one to preserve the Overhauser field even
for relatively small effective electron Zeeman splittings larger than
$A/\sqrt{N}$.  

In Sec. \ref{sec:zeno} we briefly review the quantum Zeno effect and
give the corresponding main results for the case of the Overhauser
field. We start our detailed discussion in Sec. \ref{sec:hamiltonian}
by writing down the Hamiltonian for the hyperfine interaction and by 
deriving an effective Hamiltonian for the electron-mediated
interaction. In Sec. \ref{sec:shorttime} we derive an expression for
the short-time behavior of the Overhauser field mean value. In
Secs. \ref{sec:dyson} and \ref{sec:gme} we address the long-time decay
of the Overhauser field due to the electron-mediated interaction. Some
technical details are deferred to Appendices \ref{app:dipdip} and
\ref{app:measprec}.

\section{Zeno effect}\label{sec:zeno}
The suppression of the decay of a quantum state due to frequently
repeated measurements is known as the quantum Zeno effect. The concept of the
quantum Zeno effect\cite{Misra:1977a} is almost as old as quantum mechanics
\cite{VonNeumann:1932a,Peres:1993a} and it remains one of the most
intriguing quantum effects. It has been studied intensively
from the theoretical side
\cite{Koshino:2005a,Helmer:2007a,Gambetta:2008a} and also experimental 
evidence has been found in recent years.\cite{Itano:1990a}

For a two-level system initialized to the exited state, the
survival probability $P_s$ in the exited state as a function of the
elapsed time $t$ is initially given by $P_s(t)=1-c_s t^2/\tau_s^2$,
with the constant $c_s$ and the timescale $\tau_s$ being system
dependent. A projective measurement at time $\tau_m$ resets the system 
to the excited state with probability $P_s(\tau_m)$. Repeating the 
measurement $m$ times at intervals $\tau_m\ll\tau_s$, the survival
probability is $P_{s,meas}(m\tau_m)=(1-c_s \tau_m^2/\tau_s^2)^m\approx
1-c_s m \tau_m/(\tau_s^2/\tau_m)$, for $c_sm\tau_m^2/\tau_s^2\ll 1$. The
survival probability at time $t=m\tau_m$ is thus increased due to the
frequently repeated measurements: instead of a quadratic decay on a
timescale $\tau_s$ without measurements, we have a linear decay on a
timescale $\tau_s^2/\tau_m$.  

A more complex observable such as the mean of the Overhauser field
$\langle h_z(t)\rangle=\mathrm{Tr}\{h_z\rho(t)\}$, may also
show a Zeno effect. That $\langle h_z(t)\rangle$ shows an initial
quadratic decay is, however, not obvious and actually depends
on the initial state of the nuclear spin system $\rho_I(0)$. For the
short-time behavior of $\langle h_z(t)\rangle$ we expand in a Taylor
series 
\begin{equation}\label{eq:defshorttime}
\langle h_z(t)\rangle=\langle h_z(0)\rangle +t\langle
    h_z\rangle_1 +\frac{t^2}{2}\langle h_z\rangle_2+\dots ,
\end{equation}
with $\langle h_z\rangle_n=d^n\langle
h_z(t)\rangle/dt^n|_{t=0}$. If $\langle
h_z\rangle_1=0$, the $t$-linear term vanishes and the initial decay is
quadratic in time. In Sec. \ref{sec:shorttime} we calculate the
initial dynamics of $\langle h_z(t)\rangle$ and explain the conditions
under which $\langle h_z\rangle_1=0$. We find an initial decay of the
form  \begin{equation}\label{eq:hzdecaynormzenosec}
\frac{\langle h_z(t)\rangle}{\langle
h_z(0)\rangle}=1-c \frac{t^2}{\tau_{e}^2}.
\end{equation}
The timescale $\tau_e$ and the constant $c$ are given below in
Eq. (\ref{eq:hzdecaynorm}) and Eq. (\ref{eq:c}) respectively. 

\begin{figure}
\begin{center}
\includegraphics[clip=true,width=8.5 cm]{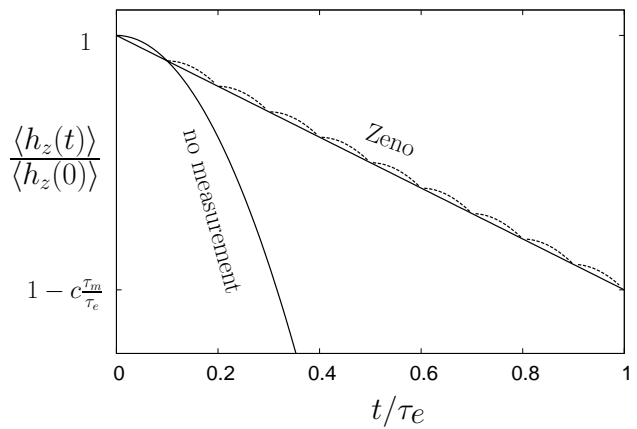}
\caption{\label{fig:zeno} Effect of projective measurements at time
  intervals $\tau_{m}=\tau_{e}/10$ on the time evolution of the 
  Overhauser field expectation value $\langle h_z(t)\rangle$. Due
  to the Zeno effect, the decay with measurements is $1-ct/\tau_{zeno}$
  rather than $1-ct^2/\tau_{e}^2$ without measurements, where
  $\tau_{zeno}=\tau_{e}^2/\tau_{m}$. The formula $1-ct/\tau_{zeno}$
  for the decay with measurement is only strictly valid at times
  $t=m\tau_m$ with $m$ being a positive integer. After the measurement
  at $t=m\tau_m$ the decay is again quadratic with time dependence
  $\langle h_z(m\tau_m)\rangle/\langle h_z(0)\rangle
  -c(t-m\tau_m)^2/\tau_{e}^2$ (broken lines).}  
\end{center}
\end{figure}

Let us now consider a sequence of repeated measurements of the
Overhauser field $h_z(t)$. In the context of quantum dots,
several proposals \cite{Stepanenko:2006a,Klauser:2006a,Giedke:2006a}
to implement such measurements have been put forward. A measurement of
$h_z$ shall be performed after a time $\tau_m$. If this measurement is
projective, i.e., if it sets all the off-diagonal elements of the
density matrix in a basis of $h_z$-eigenstates to zero (we discuss
requirements on the accuracy of the measurement in Appendix
\ref{app:measprec}), the dynamics after $\tau_m$ again follow
Eq. (\ref{eq:hzdecaynormzenosec}). Repeating the measurement at 
times $2\tau_m, 3\tau_m,$ $\dots$, leads to a change of the decay of
the Overhauser field in the same way as we described it for the
two-level system above:
\begin{equation}\label{eq:hzzeno}
\frac{\langle h_z(t)\rangle_{zeno}}{\langle h_z(0)\rangle}=1
-c \frac{t}{\tau_{zeno}},\,\,\,\tau_{zeno}=\frac{\tau_{e}^2}{\tau_m}.
\end{equation}
Instead of a quadratic decay $\propto t^2/\tau_{e}^2$ we have a linear
decay $\propto t/\tau_{zeno}$ with $\tau_{zeno}=\tau_{e}^2/\tau_m$. We
note that the expression for $\langle h_z(t)\rangle_{zeno}$ in
Eq. (\ref{eq:hzzeno}) is only strictly valid at times $m\tau_m$  with
$m$ being a positive integer. Between these times $\langle
h_z(t)\rangle$ changes according to
Eq.(\ref{eq:hzdecaynormzenosec}). The derivation of
Eq. (\ref{eq:hzzeno}) requires
$cm\tau_m^2/\tau_{e}^2=ct/\tau_{zeno}\ll 1$. Fig. \ref{fig:zeno}
shows the Zeno effect, i.e., the difference between $\langle
h_z(t)\rangle/\langle h_z(0)\rangle$ and $\langle  
h_z(t)\rangle_{zeno}/\langle h_z(0)\rangle$.

In addition to requirements on the measurement accuracy (see Appendix
\ref{app:measprec}), the results in this section rest on the following
separation of timescales:
\begin{equation}
\tau_{pm}\ll\tau_m\ll\tau_{e},\tau_x,
\end{equation}
where $\tau_{pm}$ is the time required to perform a single measurement and
$\tau_x$ the timescale up to which the short-time expansion for $\langle
h_z(t)\rangle$ is valid. In general, $\tau_x$ can be shorter 
than $\tau_e$. A specific case (fully polarized nuclear state), where the
short-time expansion has only a very limited range of validity, is
discussed in Sec. \ref{app:fullpol}. For the systems studied in
experiment, we expect $\tau_x$ to be comparable to or longer than
$\tau_e$, since the experiments performed so far show timescales for
the decay of $\langle h_z(t)\rangle$ on the order of seconds, minutes,
or in one case, even hours.\cite{Greilich:2007a} We note that it may
be a demanding task to perform the fast and precise measurements
required to obtain a Zeno effect in the present content. Still,
experimental progress in the control of the nuclear field, such as
that shown in Ref. \onlinecite{Greilich:2007a}, suggests that such
measurements may be within reach in the near future.    

We continue our discussion by deriving the effective Hamiltonian we
use both for calculating short-time dynamics and the long-time
behavior of $\langle h_z(t)\rangle$.

\section{Hamiltonian}\label{sec:hamiltonian}
We aim to describe the dynamics of many nuclear spins surrounding a
central confined electron spin in a material with an s-type conduction
band (e.g. GaAs, Si, etc.), where the dominant type of hyperfine
interaction is the Fermi contact hyperfine interaction. The electron
may be confined in many nanostructures such as nanowires, quantum 
dots or defect centers. Under the assumption that other possible sources 
of nuclear spin dynamics, such as nuclear quadrupolar coupling, are
suppressed,\cite{quadrupole} the two strongest interactions between nuclear
spins in these nanostructures are the electron-mediated interaction
(``J-coupling'' in NMR\cite{Ramsey:1952a,Slichter:1989a}) and the direct
dipole-dipole interaction. It turns out that, for a large number of
nuclei $N$ and up to magnetic fields of
a few Tesla (for GaAs), the contribution of the electron-mediated
interaction to the initial decay of the Overhauser field is dominant
(see Appendix \ref{app:dipdip}).  The Hamiltonian contains three
parts: The electron and nuclear Zeeman energies and the Fermi
contact hyperfine interaction:  
\begin{equation}\label{eq:fullhamil}
H=H_e+H_n+H_{en}=\epsilon_zS_z+\eta_z\sum_kI_k^z+\vec{S}\cdot\vec{h}.
\end{equation}
Here, the operator
\begin{equation}
\vec{h}=\sum_kA_k\vec{I}_k
\end{equation}
is the Overhauser field. Further,
$\vec{S}$ is the electron spin and $\vec{I}_k$ the 
nuclear spin at lattice site $k$ that couples with strength 
$A_k=A\nu_0|\psi(r_k)|^2$ to the electron spin, where $A=\sum_kA_k$ 
is the total hyperfine coupling constant, $\nu_0$ the volume occupied
by a single-nucleus unit cell and $\psi(r_k)$ the electron envelope wave function. 
We define the number of nuclear spins $N$ interacting with the 
electron as the number of nuclear spins within an envelope-function 
Bohr radius of the confined electron.\cite{Coish:2004a} 
The Bohr radius $a_B$ for an isotropic electron envelope is 
defined through\cite{Coish:2004a}
$\psi(r_k)=\psi(0)e^{-(r_k/a_B)^q/2}$,where $q=1$ gives a
hydrogen-like wave function and $q=2$ a Gaussian.
Finally, $\epsilon_z$ and $\eta_z$ are the electron and
nuclear Zeeman splittings, respectively (we consider a homonuclear
system). We derive an effective Hamiltonian for the
electron-mediated interaction between nuclear spins, which is valid in
a sufficiently large magnetic field. Using a standard
Schrieffer-Wolff transformation \cite{Schrieffer:1966a}
$H_{\mathrm{eff}}=e^SHe^{-S}$, in lowest order in $H_{en}$, with the transformation matrix 
$S=\sum_kA_k\left((\epsilon_z+h_z-\eta_z+A_k/2)^{-1}S_+I_k^-\right.$
 $\left. -(\epsilon_z+h_z-\eta_z-A_k/2)^{-1}S_-I_k^+\right)/2$, which
 eliminates the off-diagonal terms between electron and nuclear spins,
 we find the effective Hamiltonian $H_{\mathrm{eff}}\simeq H_0+V$ 
 (similar to Refs. \onlinecite{Shenvi:2005a,Yao:2006a,Coish:2007a}),
 where: 
\begin{eqnarray}
H_0&=&\epsilon_zS_z+\eta_z\sum_kI_k^z+S_zh_z,\\ 
V&=&\frac{1}{4(\epsilon_z-\eta_z+h_z)}
\left(\{h_-,h_+\}S_z+\frac{1}{2}[h_-,h_+]\right).\nonumber\\ \label{eq:V}
\end{eqnarray}
In Eq. (\ref{eq:V}) we have neglected terms which are suppressed by a
factor $A_k/(\epsilon_z-\eta_z+h_z)$ and the raising and lowering
operators are defined as $S_{\pm}=S_x\pm iS_y$ and similarly for
$h_{\pm}$ and $I_k^{\pm}$. The commutator $[h_-,h_+]$ is defined in
the usual way and $\{h_-,h_+\}=h_-h_++h_+h_-$ is the anti-commutator
of $h_-$ and $h_+$. We note that $H_{\mathrm{eff}}$ neglects the
transfer of spin polarization from the electron to the nuclei. 
The electron transfers an amount of angular momentum to the nuclear
system on the order $(A/\sqrt{N}\omega)^2\ll 1$ for $\omega\gg
A/\sqrt{N}$. For $\omega\sim A$ these contributions are suppressed
by a factor of $O(1/N)$ compared to the decay of $\langle h_z(t)\rangle$
under $H_{\mathrm{eff}}$. For very special initial 
states, where  $H_{\mathrm{eff}}$ leads to no dynamics, e.g., for
uniform polarization, the transfer of spin from the electron to the
nuclei is the only source of nontrivial nuclear spin dynamics and
therefore should be taken into account. We discuss one such initial
state, namely, a fully polarized nuclear system, in
Sec. \ref{app:fullpol}.
   
In the following we further replace $h_z$ in the denominator of
Eq. (\ref{eq:V}) by its initial expectation value $\langle h_z
\rangle=\mathrm{Tr}\{h_z\rho(0)\}$ and introduce the effective
electron Zeeman splitting 
\begin{equation}
\omega=\epsilon_z-\eta_z+\langle h_z
\rangle\approx \epsilon_z+\langle h_z \rangle.
\end{equation}
This replacement assumes that the initial state does not change
significantly and is valid up to corrections suppressed by
$\sigma/\omega$, compared to the dynamics under $H_{eff}$. Here
$\sigma=\sqrt{\langle h_z^2\rangle  -\langle h_z\rangle^2}$ is the
initial width of $h_z$. For an unpolarized equilibrium (infinite
temperature) nuclear spin state we have $\sigma\propto A/\sqrt{N}$,
limiting the range of validity to $\omega\gg A/\sqrt{N}$. Further
restricting our treatment to $I=1/2$ we may write $V$ as 
\begin{equation}\label{eq:V2} 
V\cong\frac{1}{2\omega}\left(S_z\sum_{\exsum{k}{l}}A_kA_lI_k^+I_l^-
  +\frac{1}{2}\sum_k A_k^2(S_z-I_k^z)\right), 
\end{equation}
where in the sum over $k$ and $l$ the terms $k=l$ are excluded. In the
next sections we will discuss the dynamics of  
the Overhauser field both at short and at long times in the regimes
where a perturbative treatment in $V$ is appropriate.

\section{Short-time expansion}\label{sec:shorttime}
With respect to the Zeno effect as discussed in Sec. \ref{sec:zeno},
our main interest lies in the short-time behavior of $\langle  
h_z(t)\rangle$ (see Eq. (\ref{eq:defshorttime})). To calculate $\langle
h_z\rangle_1$ and $\langle h_z\rangle_2$, we expand 
\begin{equation}
\langle
h_z(t)\rangle =\mathrm{Tr}\{h_z\exp{(-iHt)}\rho(0)\exp{(iHt)}\}
\end{equation} 
at short times. The first term $\langle
h_z(0)\rangle=\mathrm{Tr}\{h_z\rho(0)\}$ gives the expectation value
at time zero, while the $t$-linear term is proportional to
$\langle h_z\rangle_1=-i\mathrm{Tr}\{h_z[H,\rho(0)]\}$. Using the cyclicity  
of the trace we find that $\mathrm{Tr}\{h_z[H,\rho(0)]\}
=\mathrm{Tr}\{[\rho(0),h_z]H\}$. Writing
$\rho(0)=\rho_e(0)\otimes\rho_I(0)$ we have, for an initial nuclear
spin state $\rho_I(0)$ without transverse coherence,
$[\rho_I(0),h_z]=0$ and thus the $t$-linear term vanishes.

\begin{figure}
\begin{center}
\includegraphics[clip=true,width=8.8 cm]{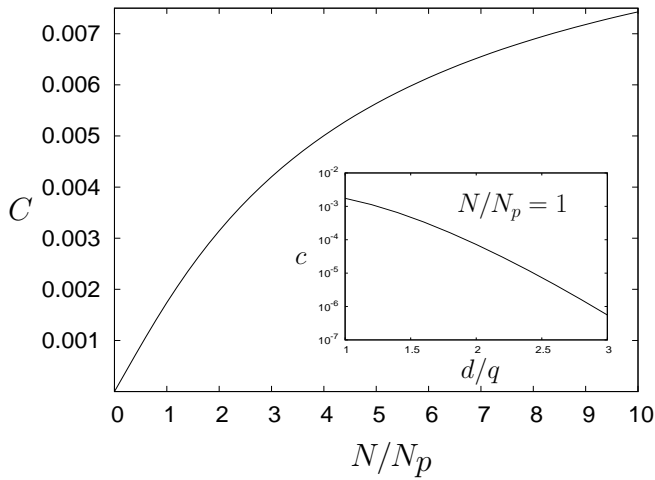}%
\caption{\label{fig:shorttime} Numerical prefactor $c$ (given in
  Eq. (\ref{eq:c})) of the
  $t^2$-term in  the decay of the Overhauser field mean value
  $\langle h_z(t)\rangle$. While $c$ turns out to be independent
  (for $N\gtrsim   100$ in the case shown according to numerical
  summation) of the number $N$ of nuclear spins within a Bohr radius of
  the electron envelope wave function, it does depend on the type of 
  structure and the initial polarization. We show the case of a 2-d
  quantum dot with a Gaussian electron envelope ($d/q=1$). The
  dependence on the initial polarization is parameterized by $N/N_p$,
  where $N_p$ is the number of nuclear spins that is polarized
  substantially (see text). Inset: dependence of $c$ on the ratio
  $d/q$ for $N/N_p=1$. We see that, e.g., for a donor impurity with a
  hydrogen-like wave function ($d/q=3$) the prefactor $c$ is more than
  three orders of magnitude smaller compared to the 2-d lateral
  quantum dot with $d/q=1$.}
\end{center}
\end{figure}

To determine the frequency of projective measurements required to 
induce a Zeno effect, we are interested in $\langle
h_z \rangle_2=-\mathrm{Tr}\{h_z[H,[H,\rho(0)]]\}$. We calculate
$\langle h_z \rangle_2$ below using the effective Hamiltonian
$H_{\mathrm{eff}}$ as derived in Sec. \ref{sec:hamiltonian}. The range
of validity is limited by higher-order terms in the effective 
Hamiltonian which are proportional to $(h_+h_-)^n/\omega^{(n+1)},
n=2,4,\dots$  These
higher-order terms give corrections to $\langle h_z\rangle_2$ which are
suppressed by a factor $(A/\sqrt{N}\omega)^n$. Thus the results for
$\langle h_z(t)\rangle$ up to $O(t^2)$ given below are valid in the
regime $\omega\gg A/\sqrt{N}$. Using that $[h_z,H_0]=0$, we may  simplify $\langle
h_z \rangle_2$ considerably and we find for an arbitrary electron spin state:
\begin{equation}
\langle h_z\rangle_2=-\frac{1}{8\omega^2}
\mathrm{Tr}_I\{h_z[\rho_I(0),h_+h_-]h_+h_-\}. 
\end{equation} 
To further simplify, we assume a product initial state of the form
\begin{eqnarray}
\rho(0)&=&\rho_e(0)\otimes\rho_I(0)=\rho_e(0)\otimes_k\rho_{I_k},\\
\rho_{I_k}&=&1/2+f_kI_k^z;\,\,\, f_k\equiv f_k(0)=2\langle
I_k^z(0)\rangle.
\end{eqnarray}
For simplicity we restrict our treatment to $I=1/2$ and thus $f_k \in
[-1,1]$. With this we find 
\begin{equation}\label{eq:g}
\langle h_z\rangle_2=-\frac{1}{4\omega^2}\sum_{kl}f_kA_k^2A_l^2
\mathrm{Tr}_I\{h_z\bigotimes_{j\neq
  k,l}(\frac{1}{2}+f_jI_j^z)(I_k^z-I_l^z)\}.
\end{equation} 
Evaluating the commutators and the trace, we find for the decay of the 
Overhauser field mean value $\langle h_z(t)\rangle$, up to corrections
of $O(t^4)$,
\begin{equation}\label{eq:hzdecay}
 \langle h_z(t)\rangle=\langle
 h_z(0)\rangle-\frac{t^2}{(8\omega)^2}\sum_{kl}A_k^2A_l^2(A_k-A_l)(f_k-f_l).  
\end{equation}
We note that both for uniform coupling constants $A_k=A/N$
and for uniform polarization $f_k=p, \forall k$, the $t^2$-term
vanishes. This is, in fact, what one would expect, since
$H_{\mathrm{eff}}$ only leads to a redistribution of polarization and
both for uniform polarization and uniform coupling constants, such a
redistribution does not affect $h_z$. Rewriting the sum
in Eq. (\ref{eq:hzdecay}) we obtain (again up to corrections of
$O(t^4)$) 
\begin{equation}\label{eq:hzdecaynorm}
\frac{\langle h_z(t)\rangle}{\langle
h_z(0)\rangle}=1-c \frac{t^2}{\tau_{e}^2},\,\,\,\tau_{e}=
\frac{N^{3/2}\omega}{A^2},
\end{equation}
with the numerical factor $c$ only depending on the distribution of
coupling constants through $\alpha_k=NA_k/A$ and the initial 
polarization distribution $f_k$ through
\begin{equation}\label{eq:c}
c=\frac{1}{32Nc_0}\sum_{kl}\alpha_k^2\alpha_l^2(\alpha_k-\alpha_l)(f_k-f_l),
\end{equation}
where $c_0=\sum_kf_k\alpha_k$. We note that, up to the factor $c$ (see
Fig. \ref{fig:shorttime}), the
timescale $\tau_{e}$ agrees with a previous rough estimate  
\cite{Klauser:2006a} for the timescale of nuclear-spin dynamics under
the electron-mediated nuclear spin interaction. In Table \ref{table:taue}
we give $\tau_{e}$ for a variety of values of the number of nuclear
spins $N$ and of $\omega=\epsilon_z-\eta_z+\langle h_z \rangle$. 
\begin{table}[!h]
\begin{tabular}{|c|c||c|c|c|c|c|} \hline
 &  &$\tau_e$ at&$\tau_e$ at&$\tau_e$ at&$\tau_e$ at&$\tau_e$ at\\
  $N$&$A/g\mu _B\sqrt{N}$& $\omega=A/\sqrt{N}$&
  100mT& 1T& 2mT& 5T \\ \hline\hline
$10^3$&49mT&3ns&6ns&60ns&119ns&297ns\\ \hline
$10^4$&16mT&29ns&188ns&2$\mu$s&4$\mu$s&9$\mu$s\\ \hline
$10^5$&49mT&292ns&6$\mu$s&60$\mu$s&119$\mu$s&297$\mu$s\\ \hline
$10^6$&1.6mT&3$\mu$s&188$\mu$s&2ms&4ms&9ms\\ \hline
\end{tabular}
\caption{\label{table:taue} This Table gives explicit values
  for the timescale $\tau_{e}$ of the $t^2$ term in the short-time
  expansion of $\langle h_z(t)\rangle$. We give $\tau_{e}$
  for various values of the number of nuclear spins $N$ and of
  $\omega=\epsilon_z-\eta_z+\langle h_z \rangle$. When
  $\omega=A/\sqrt{N}$ we are at the lower boundary of $\omega$-values
  for which the result for $\tau_{e}$ is valid. The parameters used
  are relevant for a lateral GaAs quantum dot: $A=90\mu$eV, $g=-0.4$.}
\end{table}

The coupling constants $A_k$ have a different dependence on $k$, depending  
on the dimension $d$ and the exponent $q$ in the electron envelope
wave function through\cite{Coish:2004a} $A_k=A_0e^{-(k/N)^{q/d}}$. 
For a donor impurity with
a hydrogen-like exponential wave function we have $d=3,q=1,d/q=3$,
whereas for a 2-dimensional quantum dot with a Gaussian envelope
function we have $d=2,q=2,d/q=1$. 
In Fig. \ref{fig:shorttime} we show the
constant $c$ for the case $d/q=1$ and a particular choice of the
polarization distribution. We give the dependence on $d/q$ in 
the inset of Fig. \ref{fig:shorttime}. While $c$ is independent of $N$
for $N\gtrsim 100$, it changes considerably depending on the initial
nuclear spin state, which is parameterized by the $f_k$. Since there
are neither experimental data nor theoretical calculations on the
shape of the polarization distribution, we assume for the curves in
Fig. \ref{fig:shorttime} that it has the same shape as the
distribution of coupling constants $A_k$, but with a different width,
reflected in the number of nuclear spins $N_p$ that are appreciably
polarized. The motivation for this choice is that if polarization is 
introduced into the nuclear spin system via electron-nuclear spin
flip-flops, the probability for these flip-flops is expected to be
proportional to some power of $A_k/A_0$. The degree of polarization
at the center we denote by $p\in [-1,1]$. We may thus write
$f_k=pe^{-(k/N_p)^{q/d}}$. We see in Fig. \ref{fig:shorttime} that $c$
grows monotonically with $N/N_p$, i.e., a localized polarization
distribution ($N/N_p > 1$) decays more quickly than a wide spread one
($N/N_p<1$).  

In the context of state narrowing,
\cite{Stepanenko:2006a,Klauser:2006a,Giedke:2006a} the short-time 
behavior of the width of the Overhauser field $\sigma(t)=\sqrt{\langle
  h_z^2(t)\rangle -\langle h_z(t)\rangle^2}$ is also of interest. Nuclear
spin state narrowing, i.e., the reduction of $\sigma$, extends the
electron spin decoherence time. Repeating the above calculation for
$\langle h_z^2(t)\rangle$ and using the result for $\langle 
h_z(t)\rangle$ we find ( up to corrections of $O(t^4)$) for the
variance of the Overhauser field 
\begin{equation}
\sigma^2(t)=\sigma^2(0)\left(1+c_{\sigma}\frac{t^2}{\tau_{e}^2}\right),
\end{equation}
with the range of validity $\omega \gtrsim A/\sqrt{N}$, limited by
higher-order corrections to the effective Hamiltonian as in the case
of $\langle h_z(t)\rangle$. Here, the
dimensionless constant $c_{\sigma}$ is given by 
\begin{equation}
c_{\sigma}=\frac{1}{16Nc_{\sigma0}}
\sum_{kl}\alpha_k^2\alpha_l^2(\alpha_k-\alpha_l)(f_k-f_l) 
(f_k\alpha_k+f_l\alpha_l),
\end{equation}
where $c_{\sigma0}=\sum_k\alpha_k^2(1-f_k^2)$. Taking the square-root
of $\sigma^2(t)$ and expanding it for $c_{\sigma}t^2/\tau_{e}^2\ll 1$ 
we find for the width (up to corrections of $O(t^4)$)
\begin{equation}
\sigma(t)=\sigma(0)\left(1+c_{\sigma}\frac{t^2}{2\tau_{e}^2}\right).
\end{equation}
Thus, also for the width of the Overhauser field the initial dynamics 
is quadratic in time with the same timescale as the mean. 

\subsection{Fully polarized case}\label{app:fullpol}

In this section we analyze the special case of a fully polarized
nuclear spin system, where the effective Hamiltonian derived in
Sec. \ref{sec:hamiltonian} gives no dynamics and thus the corrections
due to the transfer of polarization from the electron to the nuclei
become relevant. We thus must return to the full Hamiltonian in
Eq. (\ref{eq:fullhamil}). Using the fact that the total spin
$J_z=S_z+\sum_kI_k^z$ is a conserved quantity, we transform into a
rotating frame where the Hamiltonian takes the form\cite{Coish:2004a} 
\begin{equation}
H'=(\tilde{\epsilon}_z+h_z)S_z+\frac{1}{2}\left(h_+S_-+h_-S_+\right),
\end{equation}
with $\tilde{\epsilon}=\epsilon-\eta_z$. To have any dynamics
for a fully polarized nuclear spin system (all spins
$\ket{\uparrow}$), the initial state of the electron must be
$s_{\Downarrow}\ket{\Downarrow}+s_{\Uparrow}\ket{\Uparrow}$, with
$s_{\Downarrow}\neq 0$. Since the $\ket{\Uparrow}$ part gives no
dynamics we consider $\ket{\psi(0)}=\ket{\Downarrow;
  \uparrow\uparrow\dots\uparrow}$. At any later time we may thus write  
\begin{equation}
\ket{\psi(t)}=a(t)\ket{\psi(0)}+\sum_kb_k(t)
\ket{\Uparrow;\uparrow\uparrow\dots\uparrow 
  \downarrow_k\uparrow \dots \uparrow}, 
\end{equation}
with $a(0)=1$ and $b_k(0)=0, \forall k$. The same case was
studied in Ref. \onlinecite{Khaetskii:2003a}. However, this study
was performed from the point of view of electron spin decoherence. For
the expectation value of $\langle h_z(t)\rangle$, we find, in terms of
$a(t)$ and $b_k(t)$, 
\begin{equation}
\langle
h_z(t)\rangle=\bra{\psi(t)}h_z\ket{\psi(t)}=\frac{A}{2}-\sum_k|b_k(t)|^2A_k, 
\end{equation}
where we have used the normalization condition
$|a(t)|^2+\sum_k|b_k(t)|^2=1$. Using the time-dependent Schroedinger
equation $i\partial_t\ket{\psi(t)}=H'\ket{\psi(t)}$, we obtain the 
differential equations for $a(t)$ and $b_k(t)$:
\begin{eqnarray}
\dot{a}(t)&=&\frac{i}{4}\left(2\epsilon_z+A\right)a(t)
-\frac{i}{2}\sum_kb_k(t)A_k,\\ 
\dot{b}_k(t)&=&-\frac{iA_k}{2}a(t)
-\frac{i}{4}\left(2\epsilon_z+A-2A_k\right)b_k(t). 
\end{eqnarray}
Inserting a power-series Ansatz $a(t)=\sum_la^{(l)}t^l$ and
$b_k(t)=\sum_lb_k^{(l)}t^l$ into these equations and comparing
coefficients yields recursion relations of the form
\begin{eqnarray}
a^{(l+1)}&=&\frac{i}{4(l+1)}\left(2\epsilon_z+a\right)a^{(l)}
  -\frac{i}{2(l+1)}\sum_kb_k^{(l)}A_k, \nonumber\\
\\
b_k^{(l+1)}&=&-\frac{i A_k}{2(l+1)}a^{(l)}
-\frac{i}{4(l+1)}\left(2\epsilon_z+A-2A_k\right)b_k^{(l)}.\nonumber\\
\end{eqnarray}
Iterating these recursion relations using that $a(0)=1$ and $b_k(0)=0,
\forall k$, we find, neglecting corrections of $O(t^4)$,
\begin{equation}
\frac{\langle h_z(t)\rangle}{\langle
  h_z(0)}=1-\frac{1}{2A}\sum_kA_k^3t^2. 
\end{equation}
For the case of a 2-d quantum dot with Gaussian envelope wave function,
where we have $A_k=Ae^{-k/N}/N$, we find, evaluating $\sum_kA_k^3$ by
turning it into an integral in the continuum limit $N\gg1$, (again up
to corrections of $O(t^4)$) 
\begin{equation}
\frac{\langle h_z(t)\rangle}{\langle
  h_z(0)}=1-\frac{1}{6}\left(\frac{t}{\tau_c}\right)^2,
\end{equation}
where $\tau_c=N/A$. To obtain the range of validity for this result we
go to higher order in $t$. Again for the case of a 2-d quantum dot
with Gaussian envelope wave function we find up to $O(t^4)$, neglecting
terms that are suppressed by $O(1/N)$ in the $t^4$-term, 
\begin{equation}
\frac{\langle h_z(t)\rangle}{\langle
  h_z(0)}=1-\frac{1}{6}\left(\frac{t}{\tau_c}\right)^2
+\frac{1}{18}\left(\frac{t}{\tau_4}\right)^4.   
\end{equation}
Here, $\tau_4=2\sqrt{N}/\sqrt{A(2\epsilon_z+A)}$. This shows that in
some cases the higher order terms in the short-time expansion can have
a considerably shorter timescale. Comparing the short-time expansion
with a calculation for $\langle S_z \rangle$ in the case of uniform
coupling constants\cite{Khaetskii:2003a} suggests that the full dynamics
contain oscillations with a frequency $\propto \epsilon_z+A/2$, thus
limiting the range of validity of the short-time expansion to
$t\ll(\epsilon_z+A/2)^{-1}$. 

With this we finish our discussion of the short-time dynamics and of
the Zeno effect and move on to long-time behavior. We first show
the results of a Dyson-series expansion in Sec. \ref{sec:dyson} and
in Sec. \ref{sec:gme} we treat the problem using the generalized
master equation, showing that the Dyson-series expansion gives the
leading-order contribution in $A/\omega$.

\section{Dyson-series expansion}\label{sec:dyson}
In this section we calculate the expectation value of the Overhauser 
field $\langle h_z(t)\rangle$ in a Dyson-series expansion up to second 
order in the interaction $V$. This allows us to obtain the full time
dynamics of $\langle h_z(t)\rangle$. Since the Dyson-series expansion
is not a controlled expansion (it leads to non-secular divergences in
time at higher order), we will only see from the generalized
master equation calculation in Sec. \ref{sec:gme} that the Dyson
series result gives the correct leading order contribution in
$A/\omega$. Thus, the results in this section are expected to be valid
in the regime $\omega\gg A$.  

We transform all operators into the interaction picture by
$\tilde{\mathcal{O}}=e^{iH_0t}\mathcal{O}e^{-iH_0t}$. In the
interaction picture we have $\langle h_z(t)\rangle =
\mathrm{Tr}\{\tilde{h}_z\tilde{\rho}(t)\}$, with  $\tilde{h}_z=h_z$
since $[H_0,h_z]=0$. Expanding $\tilde{\rho}(t)$ in a Dyson series we
find \cite{Blum:1996a} 
\begin{eqnarray}
\tilde{\rho}(t)&=&\rho(0)-i\int_0^t dt'[\tilde{V}(t'),\rho(0)]\nonumber\\
& &-\int_0^tdt'\int_0^{t'}dt''[\tilde{V}(t'),
[\tilde{V}(t''),\rho(0)]]+O(\tilde{V}^3),\nonumber\\ 
\end{eqnarray}
where
\begin{equation}
\tilde{V}(t)\equiv
e^{iH_0t}Ve^{-iH_0t}=\frac{S_z}{2\omega}\sum_{\exsum{k}{l}}
e^{iS_{z}(A_k-A_l)t}I_k^+I_l^-.  
\end{equation}
We assume again the same initial state as in Sec. \ref{sec:shorttime}
and thus the term linear in $\tilde{V}$ will drop out under the trace
as it only contains off-diagonal terms. From the remaining two terms
we find 
\begin{eqnarray}\label{eq:hzdysont}
\langle h_z(t)\rangle&=&\langle h_z(0)\rangle
+\frac{1}{8 \omega^2}\sum_{\exsum{k}{l}}\frac{A_k^2A_l^2(f_k-f_l)}{A_k-A_l}\nonumber\\
& &\times \left(\cos\left[(A_k-A_l)\frac{t}{2}\right]-1\right).
\end{eqnarray}
We first verify that this result is consistent with the short-time
expansion in Sec. \ref{sec:shorttime}. For this we use that $A_k \leq
A_0\propto A/N$ and thus for times $t\ll \tau_c= N/A$ we may expand
the cosine in the above expression, recovering, to second order in $t$, 
the result in Eq. (\ref{eq:hzdecay}). For the full time dynamics we
note that the sum over cosines leads to a decay on a timescale of
$\tau_c=N/A$, since for $t>\tau_c$ the different cosines interfere
destructively. We illustrate this with an example: for a particular
choice of the initial polarization distribution ($d/q=1$ and $N_p=N$)
we may evaluate the sum in Eq. (\ref{eq:hzdysont}) in the continuum
limit and find
\begin{equation}\label{eq:hzdysondecay}
\frac{\langle h_z(t)\rangle}{\langle h_z(0)\rangle} =1-\frac{p}{8
  N}\frac{A^2}{\omega^2}g(t/\tau_c).
\end{equation}
The function $g(t)$ is explicitly given by
\begin{equation}
g(t)=\frac{1}{t^4}\left[t^4-16t^2+64t\sin\left(\frac{t}{2}\right)
  -256\sin^2\left(\frac{t}{4}\right)\right], 
\end{equation}
with $g(0)=0$ and $g(t\rightarrow \infty)=1$. We thus find a 
power-law decay on a timescale $\tau_c$ by an amount of $O(1/N)$.
Since the sum of cosines in Eq. (\ref{eq:hzdysont}) decays, the
remaining time-independent sum gives the stationary value (up to the
Poincar\'e recurrence time\cite{Fick:1990a}) 
\begin{equation}\label{eq:hzdyson}
\frac{\langle h_z\rangle_{\mathrm{stat}}}{\langle h_z(0)\rangle}=1
-\left(\frac{A}{\omega}\right)^2\frac{1}{4N^2c_0}
\sum_{\exsum{k}{l}}\frac{\alpha_k^2\alpha_l^2(f_k-f_l)}{\alpha_k-\alpha_l}. 
\end{equation}
For a system with a large number of nuclear spins $N\gg 1$ and a
sufficiently smooth polarization distribution, this stationary value differs only
by a term of $O(1/N)$ from the initial 
value, i.e., $\langle h_z\rangle_{\mathrm{stat}}/\langle
h_z(0)\rangle=1-O(1/N)$. 
\begin{figure}
\begin{center}
\includegraphics[clip=true,width=8.5 cm]{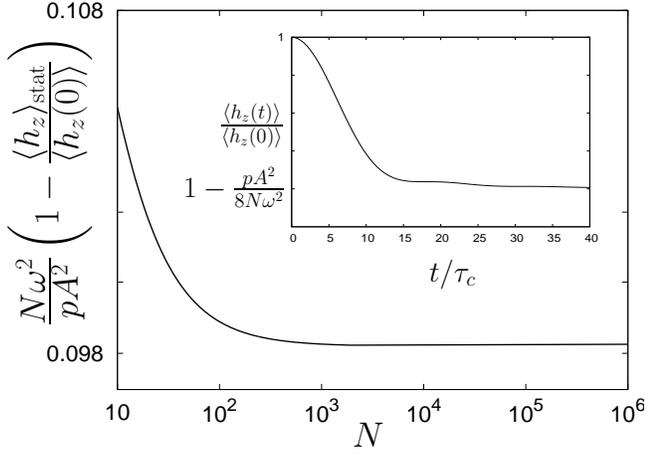}%
\caption{\label{fig:dysonN}In this figure we show the $N$-dependence of
  $1-\langle h_z\rangle_{\mathrm{stat}}/\langle h_z(0)\rangle$, i.e.,
  the part by which $\langle h_z \rangle$ decays in units of $pA^2/N\omega^2$, in the regime
  $\omega \gg A$.  This plot is for a 
  3-d defect center with a hydrogen-like electron envelope ($d/q=3$) and the initial
  polarization is parameterized by $N/N_p=0.5$ as described in
  Sec. \ref{sec:shorttime}. For this choice of polarization
  distribution the   decay is of $O(1/N)$. The inset shows the full
  time dynamics of   $\langle h_z(t)\rangle/\langle h_z(0)\rangle$ as
  given in Eq.(\ref{eq:hzdysondecay}) for $d/q=1,N \gg 1,N/N_p=1$. We
  see that the decay occurs on a timescale of $\tau_c=N/A$.} 
\end{center}
\end{figure}
This can be seen in Fig. \ref{fig:dysonN},
where we show the $N$ dependence of $1-\langle
h_z\rangle_{\mathrm{stat}}/\langle h_z(0)\rangle$, i.e., the part by
which $\langle h_z \rangle$ decays. The parameters in
Fig. \ref{fig:dysonN} are taken for a 3-d 
defect center with a hydrogen-like electron envelope ($d/q=3$) and the
initial polarization $N/N_p=0.5$ as described in
Sec. \ref{sec:shorttime}. For this choice of polarization distribution
the decay is of $O(1/N)$. We also find a $O(1/N)$ behavior for other
values of the parameters $d/q$ and $N/N_p$ and thus expect this to be
generally true for a smoothly varying initial polarization
distribution. The inset of Fig. \ref{fig:dysonN} shows the full time
dynamics of $\langle h_z(t)\rangle$ as  given in
Eq.(\ref{eq:hzdysondecay}) for $d/q=1,N\gg 1,N/N_p=1$.  

We note that the 4th order of a Dyson series expansion gives secular
terms (diverging in $t$). We thus move on to treat the long-time
behavior using a master equation approach which avoids these secular
terms and shows that the Dyson series result gives the correct 
leading-order term in $A/\omega$.

\section{Generalized master equation}\label{sec:gme}

In this section we study the decay of the Overhauser field mean value
$\langle h_z (t) \rangle$ using the Nakajima-Zwanzig generalized master
equation (GME) in a Born approximation. The results in this section
are valid in the regime $\omega\gg A$, since higher-order corrections
to the Born approximation are suppressed by a factor $(A/\omega)^2$. 

We start from the GME,\cite{Fick:1990a} which for $P_k\rho(0)=\rho(0)$ reads
\begin{equation}
P_k\dot{\rho}(t)=-iP_kLP_k\rho(t)-\int_0^tdt'P_kLe^{-iQL(t-t')}QLP_k\rho(t'),
\end{equation}
where $L=L_0+L_V$ is the Liouville superoperator defined as
$(L_0+L_V)\mathcal{O}=[H_0+V,\mathcal{O}]$. The projection superoperator
$P_k$ must preserve $\langle I_k^z(t)\rangle$ and we choose it to
have the form $P_k=\rho_e(0)\mathrm{Tr}_e\otimes
P_{dk}\bigotimes_{l\neq k}\rho_{I_{l}}(0)\mathrm{Tr}_{I_{l}}$ where
$P_{dk}$ projects onto the diagonal in the subspace of nuclear
spin $k$ and is defined as $P_{dk}\mathcal{O}=\sum_{s=\uparrow,\downarrow}
\ket{s_k}\bra{s_k}\bra{s_k}\mathcal{O}\ket{s_k}$. Further,
$Q=1-P_k$. 
In a standard Born approximation and using the same initial
conditions as above, i.e., a product state and 
no transverse coherence in the nuclear spin system, we obtain the
following integro-differential equation for $\langle I_k^z(t)\rangle$
\begin{eqnarray}\label{eq:masterborn}
\langle\dot{I}_k^z(t)\rangle=-\frac{A_k^2}{8\omega^2}\int_0^t&d\tau&
\sum_{l, l\neq k}A_l^2\cos\left[\frac{\tau}{2}(A_k-A_l)\right]\nonumber\\
& \times&(\langle I_k^z(t-\tau)\rangle-\langle I_l^z(0)\rangle).
\end{eqnarray}
The Born approximation goes to order $L_V^2$ in the expansion of the
self-energy. Higher-order corrections in $L_V$ are estimated to give
contributions to the right-hand side of Eq. (\ref{eq:masterborn}) that
are suppressed by a factor $(A/\omega)^2$. We expect the results of this
section to be valid at least for $\omega\gg A$,
although it could in principle happen that (as in the case of the
decay of $\langle S_z(t)\rangle$\cite{Coish:2004a}) the result for the
stationary value has a larger regime of validity. On the other hand it
can not be generally excluded that higher-order contributions could
dominate at sufficiently long times. Integrating
Eq. (\ref{eq:masterborn}) we find the formal solution  
\begin{eqnarray}\label{eq:masterbornintegrated}
\langle I_k^z(t)\rangle&=&\langle I_k^z(0)\rangle
-\frac{A^2}{\omega^2}\frac{\alpha_k^2}{8}\int_0^tdt'\int_0^{t'}d\tau 
\sum_{l, l\neq k}A_l^2\nonumber\\
& &\times \cos\left[\frac{\tau}{2}(A_k-A_l)\right] (\langle
I_k^z(t'-\tau)\rangle-\langle I_l^z(0)\rangle).\nonumber\\
\end{eqnarray}
This shows that
$\langle I_k^z(t)\rangle=\langle I_k^z(0)\rangle+O((A/\omega)^2)$ and we
may thus iterate this equation and replace $\langle
I_k^z(t'-\tau)\rangle$ in the integral by $\langle
I_k^z(0)\rangle$. This implies, up to corrections of $O((A/\omega)^4)$,
\begin{eqnarray}
\langle I_k^z(t)\rangle&=&\langle I_k^z(0)\rangle
-\frac{A_k^2}{16\omega^2}\sum_{l, l\neq k}A_l^2(f_k-f_l)\nonumber\\
& &\times \int_0^tdt'\int_0^{t'}d\tau\cos\left[\frac{\tau}{2}(A_k-A_l)\right].
\end{eqnarray}
Performing the integrals and summing over the $\langle
I_k^z(t)\rangle$ weighted by their coupling constants $A_k$, we recover the
Dyson series result in Eq. (\ref{eq:hzdysont}). This shows that the
Dyson series expansion gives the leading-order contribution in
$A/\omega$. 

For the analytical solution of Eq. (\ref{eq:masterborn}) in the
stationary limit we perform a Laplace transformation, 
solve the resulting equation in Laplace space, and calculate the residue
of the pole at $s=0$ which yields (up to the recurrence time)
\begin{eqnarray}\label{eq:asymdiscrete}
\langle I_k^z\rangle_{\mathrm{stat}}&=&\lim_{T\rightarrow
  \infty}\frac{1}{T} \int_0^T \langle I_k^z(t)\rangle
dt = \lim_{s\rightarrow 0}s\langle I_k^z(s)\rangle\nonumber\\
&=& \frac{1}{Z_k}\sum_{l}P_k(l) \langle I_l^z(t=0)\rangle,
\end{eqnarray}
with $Z_k=\sum_{l}P_k(l)$. We see that
$\langle I_k^z\rangle_{\mathrm{stat}}$ is determined by weighting
the neighboring $\langle I_l^z(t=0)\rangle$ with the probability
distribution $P_k(l)/Z_k$, which is explicitly given by
\begin{equation}
P_k(l)=\left\{
\begin{array}{lcr}
A_l^2/(A_k-A_l)^2&:&l\neq k,\\
2\omega^2/A_k^2&:&l=k.\\
\end{array}\right. 
\end{equation}
We point out that $\langle I_k^z\rangle_{\mathrm{stat}}$ can be either
smaller or larger than $\langle I_l^z(t=0)\rangle$ and that
$\sum_k\langle I_k^z\rangle_{\mathrm{stat}}=\sum_k\langle
I_l^z(t=0)\rangle$ since the total spin is a conserved quantity. Again
expanding the result in Eq.(\ref{eq:asymdiscrete}) to leading order in
$A/\omega$ and summing over the nuclear spins weighted by their
coupling constants $A_k$, we recover the same result found in the
Dyson series calculation in Eq. (\ref{eq:hzdyson}). Intuitively one
would expect a decay even at high fields (although a very slow one) to
a state with uniform polarization. The fact that our calculation shows
no such decay suggests that the Knight-field gradient, i.e., the
gradient in the additional effective magnetic field seen by the
nuclei, due to the presence of the electron, is strong enough to
suppress such a decay if the flip-flop terms are sufficiently
suppressed. 

As discussed in Sec. \ref{sec:dyson}, the decay to the stationary
value occurs on a timescale $\tau_c=N/A$. Performing a projective
measurement at a time $t>\tau_m$ resets the initial condition and thus
again a small decay occurs. Repeating these measurements at intervals
longer than $\tau_m$ thus allows for a decay of $\langle h_z(t)\rangle$
to zero. 

\section{Conclusion}

We have studied the dynamics of the Overhauser field
generated by the nuclear spins surrounding a bound electron. We
focused our analysis on the effect of the electron-mediated
interaction between nuclei due to the hyperfine interaction. At short
times we find a quadratic initial decay of the Overhauser field
mean value $\langle h_z(t)\rangle$ on a timescale
$\tau_e=N^{3/2}\omega/A^2$. Performing repeated strong measurements on
$h_z$ leads to a Zeno effect with the decay changing from quadratic
to linear, with a timescale that is prolonged by a factor
$\tau_{e}/\tau_m$, where $\tau_m$ is the time between consecutive
measurements. In Secs. \ref{sec:dyson} and \ref{sec:gme} we have
addressed the long-time decay of $\langle h_z(t)\rangle$ using a Dyson
series expansion and a generalized master equation approach. Both show
that $\langle h_z(t)\rangle$ only decays by a fraction of $O(1/N)$ for
a sufficiently smooth polarization distribution and large magnetic
field. It remains a subject of further study beyond the scope of this
work whether, and on what timescale, the combination of
electron-mediated interaction and direct dipole-dipole interaction may
lead to a full decay of the Overhauser field. Another interesting
question concerns the distribution of nuclear polarization within a
quantum dot or defect center and its dependence on the method that is
used to polarize the system.  

We thank G. Burkard, A. Imamo{\u g}lu, T. Meunier, K. C. Nowack,
D. Stepanenko, J.M. Taylor,  M. Trif, and in particular F.H.L. Koppens 
and L.M.K. Vandersypen for useful discussions.
We acknowledge financial support from JST ICORP, the NCCR Nanoscience
and the Swiss NSF.   

\appendix

\section{Estimation of dipole-dipole contribution}\label{app:dipdip}

In this appendix we estimate the timescale arising from the direct
secular (terms conserving $I_{z,tot}=\sum_kI_k^z$) dipole-dipole interaction in
the short-time expansion of the 
Overhauser field mean value $\langle h_z(t)\rangle$. This gives us
the range of validity of our calculation in the main text that only
took into account the electron-mediated interaction between
nuclei. Let us thus consider the situation where the external magnetic 
field is very high, such that the electron-mediated flip-flop terms
are fully suppressed. In this case the Hamiltonian has the form
$H_{dd}=H_{0,dd}+V_{dd}$, with  
\begin{eqnarray}
H_{0,dd}&=&\epsilon_zS_z+\eta_z\sum_kI_k^z+S_zh_z-
4\sum_{\exsum{k}{l}}b_{kl}I_k^zI_l^z,\nonumber\\
\\  
V_{dd}&=&\sum_{\exsum{k}{l}}b_{kl}I_k^+I_l^-.
\end{eqnarray}
Here, $b_{kl}=\gamma_I^2(3\cos^2(\theta_{kl})-1)/r_{kl}^3$, with $\theta_{kl}$ being the angle
between a vector from nucleus $k$ to nucleus $l$ and the $z$-axis and
$r_{kl}$ being the distance between the two
nuclei.\cite{Slichter:1989a} Further, $\gamma_I$ is the nuclear gyromagnetic
ratio. For the short-time expansion, only
the off-diagonal terms are relevant, since
$[h_z,H_0]=[\rho(0),H_0]=0$. These off-diagonal terms in the case of
the electron-mediated interaction are  
$S_z\sum_{\exsum{k}{l}}A_kA_lI_k^+I_l^-/2\omega$ (see Eq. (\ref{eq:V2})). Replacing
$A_kA_l/2\omega$ by $b_{kl}$ in the result for the short-time
expansion in Eq.(\ref{eq:hzdecay}) and also taking into account the
factor of $1/4$ that comes from $S_z^2$ in the electron-mediated case
we find 
\begin{equation}\label{eq:hzdecaydipdip}
 \langle h_z(t)\rangle_{\mathrm{dip-dip}}=\langle
 h_z(0)\rangle-\frac{t^2}{4}\sum_{kl}b_{kl}^2(A_k-A_l)(f_k-f_l).  
\end{equation}
To estimate, we restrict the sum to nearest neighbors as the $b_{kl}$
fall off with the third power of the distance between the two
nuclei. Assuming $f_k=(A_k/A_0)^{N/N_p}$ we find up to corrections of
$O(t^4)$  
\begin{equation}
\frac{\langle h_z(t)\rangle_{\mathrm{dip-dip}}}{\langle
h_z(0)\rangle}\approx1-\frac{t^2}{\tau_{d}^2},
\,\,\,\tau_{d}=\frac{\sqrt{N_pN}}{b}, 
\end{equation}
with $b$ being the nearest-neighbor dipole-dipole coupling. For GaAs
we have $b\sim 10^2 s^{-1}$ (with $\gamma_I\approx10$
MHz/T \cite{Mao:1995a}). For $NN_p\gg 1$ we have
$\tau_d\gg10^{-2}s$. In the magnetic field range shown in Table  
\ref{table:taue} we thus have $\tau_d\gg \tau_e/\sqrt{c}$, which
justifies neglecting the direct dipole-dipole coupling in the
short-time expansion.

\section{Measurement accuracy}\label{app:measprec}
The description of the Zeno effect in Sec. \ref{sec:zeno} relied on
the assumption that the measurements on $h_z$ set all off-diagonal
elements of the density matrix to zero. This assumption requires on
one hand a perfect measurement accuracy for $h_z$ (we discuss
deviations from that below), but on the other hand it also requires
the $h_z$-eigenstates to be non-degenerate. For non-degenerate $h_z$
eigenstates a measurement of $h_z$ fully determines the
polarization distribution $f_k$ and we may thus write $\rho_I$ after
the measurement again as a direct product with
$\rho_{I_k}(\tau_m)=1/2+f_k(\tau_m)I_k^z$. After the measurement, we
thus again have the same time evolution for $\langle h_z(t)\rangle$ as
given in Eq. (\ref{eq:hzdecay}), but with $f_k$ replaced by
$f_k(\tau_m)$. Iterating Eq. (\ref{eq:hzdecay}) for the case of $m$
consecutive measurements at intervals $\tau_m$ one obtains
Eq. (\ref{eq:hzzeno}).  

Instead of the idealized assumption of a projective measurement we now
allow for imperfect measurements. To describe these
measurements we use a so-called POVM (positive operator valued
measure).\cite{Peres:1993a} In a general POVM measurement the density
matrix changes according to 
\cite{Peres:1993a} 
\begin{equation}\label{eq:rhoprime}
\rho \rightarrow \rho'=\int\sqrt{F_y}\rho\sqrt{F_y}dy,
\end{equation}
when averaging over all possible measurement outcomes $y$. The
probability to measure outcome $y$ is given by $P(y)=\mathrm{Tr}\{\rho
F_y\}$ and the condition $\int dyF_y=1$ ensures that the
probabilities sum to unity. We consider the nuclear density matrix
$\rho_I$ in a basis of $h_z$ eigenstates $\ket{n}$ with
$h_z\ket{n}=h_z^n\ket{n}$. We denote the matrix elements of $\rho_I$
by $\rho_I(n,m)=\bra{n}\rho_I\ket{m}$. For the following description
we assume that the diagonal of the nuclear spin density matrix before
the measurement is Gaussian distributed around its mean value $\langle 
h_z\rangle$ with a width $\sigma$, i.e., 
\begin{equation}
\rho_I(n,n)=\frac{1}{\sqrt{2 \pi}\sigma}
\exp\left[-\frac{(h_z^n-\langle h_z\rangle)^2}{2\sigma^2}\right].
\end{equation}
For an unpolarized equilibrium (infinite temperature) state, the width is $\sigma\propto
A/\sqrt{N}$. Here, $\sigma$ can take any value. Let us now consider a
measurement that determines the value of $h_z$ up to an accuracy
$\eta$ with a Gaussian lineshape. We refer to $\eta$ as the
measurement accuracy. If the outcome of the measurement is
$\langle h_z\rangle+y$, the diagonal of the nuclear spin density
matrix after a measurement has the form   
\begin{equation}
\rho'_I(n,n;y)=\frac{1}{\sqrt{2\pi}\eta}\exp\left[-\frac{(h_z^n-\langle
    h_z\rangle -y)^2}{2\eta^2}\right].
\end{equation}
Since we aim to describe measurements that at least partially project
the nuclear spin state, we have $\eta<\sigma$. The POVM that describes
such a measurement is given by
\begin{equation}
F_y=\sum_n f(n,y)\ket{n}\bra{n},
\end{equation}
with
\begin{eqnarray}
f(n,y)&=&\frac{\sigma}{\eta\sqrt{2\pi(\sigma^2-\eta^2)}}
\exp\left[-\frac{(h_z^n-\langle h_z\rangle -y)^2}{2\eta^2}\right] \nonumber\\
  & &\times\exp\left[-\frac{(h_z^n-\langle
      h_z\rangle)^2}{2\sigma^2}-\frac{y^2}{2(\sigma^2-\eta^2)}\right].  
\end{eqnarray} 
We note that for $\eta \ll \sigma$ we have
$f(n,y)\approx\exp(-(h_z^n-\langle h_z\rangle
-y)^2/2\eta^2)/\sqrt{2\pi}\eta$.   
With $f(n,y)$, the operators $F_y$ are fully determined and it is
straightforward to calculate the probability for
obtaining the measurement result $\langle h_z\rangle+y$
\begin{equation}
P(y)=\frac{1}{\sqrt{2\pi(\sigma^2-\eta^2)}}
  \exp\left[-\frac{y^2}{2(\sigma^2-\eta^2)}\right].
\end{equation}
Clearly, the probabilities add up to one ($\int P(y) dy=1$) as they
 should. Also, when weighting the $\rho'_I(n,n;y)$ with their
probabilities for occurring, we find
$\int\rho'_I(n,n;y)P(y)dy=\rho_I(n,n)$. 
Using Eq. (\ref{eq:rhoprime}) we thus find for the matrix elements after a measurement, when
averaging over all possible measurement outcomes 
\begin{equation}
\rho_I(n,m)\rightarrow \rho'_I(n,m)=\rho_I(n,m)\int\sqrt{f(n,y)f(m,y)}dy,
\end{equation}
with (for $\eta\ll\sigma$)
\begin{equation}
f(n,y)\approx \frac{1}{\sqrt{2\pi}\eta}
\exp\left[-\frac{(h_z^n-h_{z0}-y)^2}{2\eta^2}\right]. 
\end{equation}
Again, for $\eta\ll\sigma$, we thus have
\begin{equation}
\rho'_I(n,m)=\rho_I(n,m)\exp\left[-\frac{(h_z^n-h_z^m)^2}{8 \eta^2}\right].
\end{equation}
To reduce the off-diagonal elements, the measurement accuracy must
be better than the difference in eigenvalues. In the limit $\eta
\rightarrow 0$ a projective measurement is recovered, which sets all
off-diagonal elements to zero. Up to $t^2$ in the short-time 
expansion, only off-diagonal elements between states that differ at
most by two flip-flops can become non-zero. Thus, to have at least a
partial Zeno effect,\cite{Koshino:2005a} resulting from the
off-diagonal elements being partially reduced, the requirement on the
measurement accuracy is $\eta \lesssim h_z^n-h_z^m$ with
$\ket{n}=I_k^+I_l^-I_p^+I_q^-\ket{m}$. For coupling constants
$A_k=Ae^{-k/N}/N$, we have typically $h_z^n-h_z^m\propto A/N^{3/2}$.  
Besides destroying the off-diagonal elements of $\rho_I$ through a
measurement, there are also ``natural'' dephasing mechanisms, such as
inhomogeneous quadrupolar splittings, electron-phonon coupling, or
spin-lattice relaxation, that can lead to a reduction of the
off-diagonal elements of $\rho_I$.

\bibliography{nucleardynamics}

\begin{thebibliography}{68}
\bibitem[{dil()}]{diluting}
\bibinfo{note}{{Diluting reduces the dipolar coupling strongly, as it decreases
  with the third power of the distance between the nuclei. In contrast, the
  hyperfine coupling is proportional to the density of nuclei and thus for a
  one-dimensional (two-dimensional) system only decreases with the first
  (second) power of the distance between the nuclei.}}
\end{thebibliography}
\end{document}